\documentstyle[12pt]{article}
\newcommand{\dd}{\mbox{\rm d}} 
\newcommand{\DD}{\mbox{\rm D}}
\newcommand{\nnn}{\noindent} 
\newcommand{\p}{\partial}
\newcommand{\be}{\begin{equation}} 
\newcommand{\bear}{\begin{eqnarray}}
\newcommand{\ee}{\end{equation}}
\newcommand{\eear}{\end{eqnarray}}
\newcommand{\bi}{\bibitem} 

\newcommand{\vs}{\vspace}

\newcommand{\AmS}{{\protect\the\textfont2
  A\kern-.1667em\lower.5ex\hbox{M}\kern-.125emS}}

\begin{document}

\begin{center}
{\LARGE \bf Fock-Schwinger proper time formalism for $p$-branes}\footnote{
\small Work supported by the Slovenian Ministry of Science
        and Technology under Contract J1-7455-0106-96}

\vs{6mm}

Matej Pav\v si\v c

\vs{3mm}

       Jo\v zef Stefan Institute, 
        University of Ljubljana, 
        Jamova 39, 1000 Ljubljana, Slovenia\footnote{
        E-mail: MATEJ.PAVSIC@IJS.SI}
             
\vs{1cm}

ABSTRACT

\end{center}

The theory of the usual, constrained $p$-branes is embedded into a larger
theory in which there is no constraints. In the latter theory the
Fock-Schwinger proper time formalism is extended from point-particles
to $p$-branes which can be considered as a points in an infinite
dimensional space ${\cal M}$. The quantization appears to be straightforward
and elegant. The conventional $p$-brane states are particular stationary
solutions to the functional Schr\" odinger equation which describes
the evolution of a membrane's state with respect to the invariant
evolution parameter $\tau$. It is also shown that states of a lower
dimensional $p$-brane can be considered as particular states of a
higher dimensional $p$-brane.

\vs{1cm}

\section{Introduction}

Relativistic $p$-branes are being intensively studied
nowadays \cite{1}. A very interesting mathematical structure is being revealed
and many are convinced that, in one way or another, it will find its
place in physics. A consistent quantum theory of a $p$-brane can
be formulated in the embedding space of a certain characteristic dimension
(e.g. 26 for bosonic strings). The problem then occurs how to compactify
all those extra dimensions to the observed 4 dimensions. We prefer to
adopt a different view and assume that a 3-brane sweeping a 4-dimensional
worldsheet $V_4$ in an embedding space $V_N$ (where N is determined by
consistency conditions) already represents a spacetime \cite{2,2a}.
There is no need
to compactify the dimensions of $V_N$. But, unfortunately, quantization
of such a higher dimensional extended object is extremely involved
because of the presence of the constraints due to the reparametrization
invariance. As a way to avoid such difficulties I proposed to consider
the so called unconstrained $p$-branes \cite{3}. The latter
objects I shall often
call simply "membranes" (their dimensionality not being restricted to 2).
Arbitrary deformations of a membrane are allowed and there is no
constraints. Quantization of such a system is straightforward, and
the conventional $p$-brane states occur as stationary solutions to the
Schr\" odinger equation.

\section{The classical dynamics of unconstrained membranes}

We shall first consider the classical dynamics of an $n$-dimensional
membrane ${\cal V}_n$ described by the variables $X^{\mu} (\tau, \xi^a)$,
$\mu = 1,2,...,N$ ; $a = 1,2,...,n$  which denote position of ${\cal V}_n$
in $V_N$. We assume the following action which is an extension of the
Stueckelberg action from point particles to membranes:
\be
  I = {{\kappa} \over 2} \int {\dd} \tau \, {\dd}^n \xi \, \sqrt{|f|} \,
    \biggl[ {1 \over {\Lambda}} \, g_{\mu \nu} ({\dot X}^{\mu} + 
    {\Lambda}^a \, \p_a X^{\mu} )
    ({\dot X}^{\nu} + {\Lambda}^b \p_b X^{\nu})
    + \Lambda  \biggr]
\label{1}
\ee

\nnn where $f \equiv {\rm det} f_{ab}$ is the determinant of the induced
metric $f_{ab} \equiv \p_a X^{\mu} \p_b X_{\mu}$ on ${\cal V}_n$,
$\kappa$ a constant and $g_{\mu \nu}$ the metric of $V_N$.
Variation of the action in our approach is to be performed solely with
respect to $X^{\mu}$, while $\Lambda$ and $\Lambda^a$ are assumed to be
fixed background fields on ${\cal V}_n$. In the conventional approach
$\Lambda$, $\Lambda^a$ are Lagrange multipliers leading to the
$p$-brane constraints and the action (1) is equivalent to the Dirac-Nambu-Goto
action; fixing of $\Lambda$, $\Lambda^a$ then fixes a gauge. On the
contrary, in our approach $\Lambda$, $\Lambda^a$ are not Lagrange multipliers
at all, they are fixed from the very beginning: a physical consequence is
that a membrane is arbitrarily deformable, and its tension is not
necessarily a constant, but depends on a solution of the equations
of motion \cite{2a,3}.

The quantities $X^{\mu} (\tau, \xi)$ are independent dynamical variables,
and there is no constraints. Those $n$ components among $X^{\mu} (\tau, \xi)$,
roughly speaking, which  are redundant in the conventional approach
and must be determined by choice of a gauge, are not redundant at all in
our approach: they are necessary to describe deformations of membrane
\cite{2a,3}.

The equation of motion derived from (\ref{1}) gives after contraction with
$\p^c X^{\mu}$ and rearrangement of the terms:
\be     
     {{\dd} \over {{\dd} \tau}} (p_{\mu} \, \p_c X^{\mu}) +
     \p_a (\Lambda^a \, p_{\mu} \p_c X^{\mu})  
     = {1 \over 2} \,
     \p_c \Lambda \, {{\sqrt{|f|}} \over {\kappa}}
     \left ( {{p^2} \over {|f|}} - \kappa^2 \right )
\label{2}
\ee

\nnn where $p_{\mu} =
(\kappa \sqrt{|f|}/\Lambda)({\dot X}_{\mu} + {\Lambda}^a \p_a X_{\mu})$ is
the canonical momentum. Eq.(\ref{2}) admits a solution satisfying
\be
    p_{\mu} \p_c X^{\mu} = 0
\label{3}
\ee
\be
           p^2 - |f|\, \kappa^2 = 0
\label{4}
\ee

\nnn which are just the $p$-brane constraints. Conventional $p$-branes,
satisfying the minimal surface equation are among possible
solutions to our dynamical system. When $\Lambda^a = 0$, Eq.(\ref{3}) says
that ${\dot X}^{\mu} \p_a X_{\mu} = 0$ whichs means that the velocity
${\dot X}^{\mu}$ is perpendicular to the membrane. But in ge\-neral,
$p_{\mu} \p_c X^{\mu} \neq 0$, and the velocity has non zero tangent
component to the membrane, so that different parts of the membrane move
relative to each other. Such a membrane is then a wiggly membrane \cite{4}.

Different functions $X^{\mu} (\xi)$ and ${X'}^{\mu} (\xi)$ which happen to
describe the same surface $V_n$ we can interpret

\ (i) {\it passively} (as representing the same membrane ${\cal V}_n$
in different parametrizations of $\xi^a$),

(ii) {\it actively} (as representing two physically distinct, deformed,
membranes ${\cal V}_n$ and ${\cal V'}_n$, respectively).

In order to better understand this, just imagine a rubber sheet
spanning a surface $V_2$. It may happen that two physically different
configurations ${\cal V}_2$, ${\cal V}_2^{'}$ of the deformed sheet
span the same surface $V_2$.

\section{The membrane space}

It is convenient to introduce the concept of {\it membrane space}
${\cal M}$ the points of which are unconstrained membranes ${\cal V}_n$,
parametrized by coordinates $X^{\mu} (\xi) \equiv X^{\mu(\xi)}$. {\it The
distance} is defined by
\be
    {\dd} {\ell}^2 = {\rho}_{\mu(\xi ) \nu(\zeta)} \, 
    {\dd} X^{\mu (\xi)} \, {\dd} X^{\nu (\zeta)} =
    {\dd} X^{\mu (\xi)} {\dd} X_{\mu(\xi)}
\label{5}
\ee

\nnn where the metric is
\be
       {\rho}_{\mu(\xi) \nu (\zeta)} = {{\kappa \sqrt{|f|}}\over
       {\Lambda}} \, g_{\mu \nu} \,
       \delta (\xi - \zeta)
\label{6}
\ee

\nnn In Eq.(\ref{5}) we adopt the convention of summation over repeated indices,
such as $\mu, \, \nu$ and integration over the repeated continuous
indices, such as $\xi, \, \xi'$.

The tensor calculus in ${\cal M}$ \cite{2a} is a straightforward
generalization of the tensor calculus in a finite dimensional space.
From the metric (\ref{6}) and its inverse
\be
   \rho^{\mu (\xi) \nu (\xi')}
  = {{\Lambda} \over {\kappa \sqrt{|f|}}} g^{\mu \nu} \, \delta (\xi - \xi')
\label{7}
\ee
  
\nnn we can construct the affinity $\Gamma_{\alpha(\xi') \beta(\xi'')}^{\mu
(\xi)}$ and define the covariant (functional) derivative ${\rm D}/
{\rm D} X^{\mu(\xi)} \equiv {\rm D}_{\mu (\xi)}$ in ${\cal M}$.
When acting on a scalar
functional $A[X(\xi)]$ the covariant derivative coincides with the
partial functional derivative $\delta/\delta X^{\mu} (\xi) \equiv
\p/\p X^{\mu(\xi)} \equiv \p_{\mu(\xi)}$. But in general, a geometric
object in ${\cal M}$ is a tensor of arbitrary rank,
${A^{\mu (\xi)}}_{\nu (\xi')...}$ which is a function of $X^{\mu (\xi)}$
(i.e., a functional of $X^{\mu} (\xi)$).

The invariant measure in ${\cal M}$ (i.e., the volume element) is
\be
    {\cal D} X = {\left ({\rm Det} \rho_{\mu(\xi) \nu(\xi')} \right) }^{1/2} 
    \prod_{\mu, \xi} {\dd} X^{\mu} (\xi) 
     = \prod_{\mu, \xi}
    {\left ( {{\kappa \sqrt{|f|}}\over {\Lambda}} \right )}^{1/2}
    {\dd} X^{\mu} (\xi)
\label{8}
\ee

The action (\ref{1}) can be written in the compact notation of 
${\cal M}$ space as
\be
     I = {1 \over 2} \int {\dd} \tau \, \biggl[ {1 \over {\Lambda}}
     \rho_{\mu (\xi) \nu (\xi')} ({\dot X}^{\mu (\xi)} + \Lambda^a
     \p_a X^{\mu (\xi)}) 
     ({\dot X}^{\nu (\xi')} + \Lambda^b \p_b
     X^{\nu (\xi')}) + K \biggr] 
\label{9}
\ee

\nnn where $K \equiv \int {\dd}^n \xi \, \sqrt{|f|} \, \Lambda \kappa $.
When $\Lambda^a = 0$ Eq.(\ref{9}) has the same form as the well known
Stueckelberg point-particle action \cite{5}.
However, the space in which the dynamics takes
place is now the infinite dimensional membrane space ${\cal M}$, and
the evolution parameter $\tau$ is essentially the proper time in ${\cal M}$.
The spacetime, which is the arena for the dynamics of a Stueckelberg point
particle, is just a subspace of ${\cal M}$. The action (\ref{9}) is a
generalization of the Stueckelberg action and it describes membranes of
arbitrary dimensions $n$, including point-particles, when $n=0$.      

{\it The Hamiltonian} belonging to (\ref{1}) is
\be
     H = \int {\dd}^n \xi \, \biggl[ \sqrt{|f|} \, 
     {\Lambda \over {2 \kappa}} \left ( {{p^{\mu} p_{\mu}} \over {|f|}}
     - \kappa^2 \right ) 
     - \, \Lambda^a \, \p_a X^{\mu} p_{\mu} \biggr]
\label{10}
\ee

\nnn and can be written compactly as
\be
      H = {1 \over 2} (p^{\mu (\xi)} p_{\mu (\xi)} - K ) -
      \Lambda^a \p_a X^{\mu (\xi)} p_{\mu (\xi)}
\label{11}
\ee

\nnn where and $p_{\mu (\xi)} \equiv p_{\mu} (\xi)$. The canonical and the
Hamilton-Jacobi theory for unconstrained membranes can be straightforwardly
developed \cite{6}.

\section{The quantum theory of unconstrained membranes}

Quantization of the theory is performed by considering $X^{\mu} (\xi)$
and $p_{\mu} (\xi)$ as operators satisfying the commutation relations
\be
    [X^{\mu} (\xi), \, p_{\nu} (\xi')] = i\, {\delta^{\mu}}_{\nu}\, 
    \delta (\xi - \xi')
\label{12}
\ee
\be
     [X^{\mu} (\xi), \, X^{\nu} (\xi')] = 0 \; \; , \quad
     [p_{\mu} (\xi), \, p_{\nu} (\xi')] = 0
\label{13}
\ee

\nnn In the coordinate representation $X^{\mu} (\xi)$ are diagonal. The
momentum operator is
$p_{\mu} (\xi) = - i \delta/\delta X^{\mu} (\xi)$, when acting on a scalar
functional; otherwise it is given by the covariant derivative
$p_{\mu} (\xi) \equiv p_{\mu (\xi)} = - i {\DD}_{\mu (\xi)}$.
A state is represented by a $\tau$-dependent wave functional
$\psi [\tau, X^{\mu} (\xi)]$ sa\-tisfying the Schr\" odinger equation
\be        
     i \hbar \, {{\p \psi} \over {\p \tau}} = H \psi
\label{14}
\ee

\nnn where $H$ is given by (\ref{11}) with $p_{\mu (\xi)} =
- i {\cal D}_{\mu (\xi)}$. Eq.(\ref{14}) is a generalization of
the well known point-particle relativistic wave equation with
proper time \cite{5}.

The wave functional $\psi$ is normalized according to \cite{5}
\be
     \int {\cal D} X \, \psi^* \psi = 1
\label{15}
\ee

\nnn where ${\cal D} X$ is given in (\ref{8}). Eq.(\ref{15}) is
a straightforward extension of the corresponding relation
$\int {\dd}^4 x \, \psi^* \psi = 1$ for the unconstraint point particle
in Minkowski spacetime \cite{5,5a}. It is important to stress that,
since (\ref{15}) is satisfied at any $\tau$, the evolution operator
$U$ which brings $\psi (\tau) \rightarrow \psi (\tau') = U \psi (\tau)$ is
{\it unitary}, and no ne\-gative norm states occur in such a theory.

{\it Stationary solutions}  to the Schr\" odinger equation (\ref{14}) are
given by
\be
   \psi[\tau, X^{\mu}(\xi)] = e^{-iE \tau} \phi[X^{\mu}(\xi)]
\label{16}
\ee
   
\nnn where $E$ is a constant of motion, and satisfy
\be
     \left ( -\, {1 \over 2} {\DD}^{\mu (\xi)} {\DD}_{\mu (\xi)} +
     i\, \Lambda^a \p_a X^{\mu (\xi)} {\DD}_{\mu (\xi)} - {1\over 2} K
     \right ) \phi 
     = E \, \phi
\label{17}
\ee

\nnn For a real $\phi$ Eq.(\ref{17}) splits into
\be 
   \left ( -\, {1 \over 2} {\DD}^{\mu (\xi)} {\DD}_{\mu (\xi)} -
     {1\over 2} K - E \right ) \phi = 0
\label{18}
\ee

\be
    \Lambda^a \p_a X^{\mu (\xi)} {\DD}_{\mu (\xi)} \phi = 0
\label{19}
\ee

\nnn The latter equations are satisfied by the well known $p$-brane
constraints acting on a state $\phi$. We see that our Schr\" odinger
equation (\ref{14}) contains the conventional $p$-brane states as
particular solutions.

{\it Wave packets} of membrane's states can also be studied with our
approach, at least in a simple case when the centre of the wave packet
has linear dependence on $\tau$ \cite{2a}. This corresponds to the case of
a null string \cite{7}.

{\it Reduction of membrane's dimension} can be performed by considering
a special, limiting, shape of the wave packet, such that one or more
of membranes dimensions disappear. A wave packet which formally describes
a state of an $n$-dimensional membrane, effectively describes  a state
of an $(n-1)$-dimensional (or lower dimensional) membrane. This
principle holds also for statio\-nary states among which there are the
conventional $p$-brane states.

This can be shown as follows. Suppose that $\psi[\tau, X^{\mu} (\xi^a)]$
is a wave functional of an $n$-dimensional membrane. Suppose now that it
sa\-tisfies the relation
\be
     {{\delta \psi} \over {\delta X^{\mu} (\xi^{a_0}, \xi^i)}} = 
     \delta (\xi^{a_0} - \xi_{\Sigma}^{a_0} ) (\p_{a_0} X^{\mu} 
     \p_{a_0} X_{\mu})^{1/2}
     {{\delta \psi} \over {\delta X^{\mu} 
     (\xi_{\Sigma}^{a_0}, \xi^i)}}
\label{20}
\ee

\nnn where $\xi^{a_0}$ is one of the coordinates $\xi^a$, $a =1,2,...,n$
and $\xi_{\Sigma}^{a_0}$ is its fixed value. Then $X^{\mu} (\xi_{\Sigma}
^{a_0}, \xi^i) \equiv X^{\mu} (\xi^i), \quad i \neq a_0$ represents an
$n-1$ dimensional membrane ${\cal V}_{n-1}$, and we have
\bear
     {\DD}^{\mu (\xi)}{\DD}_{\mu (\xi)} \psi & &= \int {\dd}^n \xi \,
    {{\Lambda} \over {\kappa \sqrt{|f|}}} \, g^{\mu \nu} {{{\DD}^2 \psi}
    \over
    {{\DD} X^{\mu} (\xi) {\DD} X^{\nu} (\xi)}} \nonumber \\
    & & = \int {\dd}^{n-1} \xi
    {{\Lambda} \over {\kappa \sqrt{|{\bar f}|}}} \, g^{\mu \nu} 
    {{{\DD}^2 \psi} \over
    {{\DD} X^{\mu} (\xi^i) {\DD} X^{\nu} (\xi^i)}}
     = {\bar {\DD}}^{\mu(\xi^i)} {\bar {\DD}}_{\mu (\xi^i)} \psi
\label{21}
\eear

\nnn Here ${\bar f} \equiv {\rm det} \, {\bar f}_{ij} = f/\p_{a_0} X^{\mu}
\, \p_{a_0} X_{\mu}$ is the determinant of the induced metric
${\bar f}_{ij} \equiv \p_i X^{\mu} \, \p_j X_{\mu}$ on ${\cal V}_{n-1}$.

Similarly
\be
   \Lambda^a \p_a X^{\mu (\xi)} {\DD}_{\mu (\xi)} \psi = \Lambda^i
   \p_i X^{\mu (\xi^i)} {\DD}_{\mu (\xi^i)} \psi
\label{22}
\ee

\nnn The expression (\ref{21}) and (\ref{22}) enter the Hamiltonian
(\ref{11}). The Schr\" odinger equation (\ref{14}) which is formally
an equation for a state $\psi[\tau, X^{\mu} (\xi^a)]$ of a membrane
${\cal V}_n$ reduces to the equation of a state $\psi[\tau, X^{\mu} 
(\xi^i)]$ of a membrane ${\cal V}_{n-1}$.  This process can be
continued from $n-1$ to $n-2$, etc.

\section{Conclusion}

We embedded the theory o $p$-branes into a larger theory in which there
is no constrained. Conventional $p$-brane states are particular stationary
solutions to the covariant Schr\" odinger equation in which an invariant
evolution para\-meter $\tau$ takes place. We have generalized the
Fock-Stueckelberg-Schwinger proper time formalism from point-particles to
membranes. We adopt the interpretation that such a generalized theory
(without constraints) has its physical content, and is not considered
merely as a convenient mathematical tool.

Very useful is the concept of the membrane space ${\cal M}$ which
enables us to formulate the membrane's dynamics in essentially the same
manner as the dynamics of the relativistic Stueckelberg point-particle.
Construction of $p$-brane's field theory (to be presented elsewhere)
seems to emerge uniquely and straightforwardly from this approach.
Finally let us observe that the usual $p$-brane theories (with constraints)
have not yet been fully confronted with experiments, therefore it makes
sense to consider an enlarged theory. Fermions can also be considered
by extending the formalism to include the Grassmann coordinates.

\end{document}